\documentclass{aa}  
\usepackage{natbib}

\usepackage{graphicx}
\usepackage{subfigure} 
\usepackage{txfonts}
\usepackage[pdftex]{hyperref}

\begin{document} 

   \title{Infall of galaxies onto groups}

   \author{M. V. Santucho
          \inst{1,2},
          M. L. Ceccarelli\inst{1,2}
           \and
          D. G. Lambas\inst{1,2}
          }

   \institute{$^1$Instituto de Astronomía Teórica y Experimental (IATE), CONICET-UNC, Laprida 854, X5000BGR, Córdoba, Argentina\\
              \email{santucho@oac.unc.edu.ar}\\
             $^2$Observatorio Astronómico de Córdoba (OAC), Universidad Nacional de Córdoba (UNC), Córdoba, Argentina. 
         }

   \date{Received ; accepted }

\abstract{
Growth of the structure in the Universe manifest as accretion flows of galaxies onto groups and clusters.
Thus, the present day properties of groups and their member galaxies are influenced by the characteristics of this continuous infall pattern.
Several works both theoretical, in numerical simulations, and in observations, study this process and provide useful steps for a better understanding of galaxy systems and their evolution.
}{
We aim at exploring the streaming flow of galaxies onto groups using observational peculiar velocity data.
The effects of distance uncertainties are also analyzed as well as the relation between the infall pattern and
group and environment properties.
}{
This work deals with analysis of peculiar velocity data and their projection on the 
direction to group centers, to determine the mean galaxy infall flow. We applied
this analysis to the galaxies and groups extracted from the Cosmicflows--3 catalog.
We also use mock catalogs derived from numerical simulations to explore the effects 
of distance uncertainties on the derivation of the galaxy velocity flow onto groups.
}{
We determine the infalling velocity field onto galaxy groups with cz $<$ 0.033 using peculiar velocity data.
We measure the mean infall velocity onto group samples of different mass range, and also explore the impact of the
environment where the group reside. 
Well beyond the group virial radius, the surrounding large–scale galaxy overdensity may impose additional infalling streaming amplitudes in the
range 200 to 400 km/s. 
Also, we find that groups in samples with a well controlled galaxy density environment show an increasing infalling velocity amplitude
with group mass, consistent with the predictions of the linear model.
These results from observational data are in excellent agreement with those derived from the mock catalogs.
}{} 


   \maketitle
%
%
\section{Introduction}
In the nearby Universe, galaxy peculiar velocities manifest the evolution of the 
large-scale structure. This growth of structure, within hierarchical clustering scenarios, 
cause the increase of the masses of galaxy groups and clusters through the continuous accretion of smaller systems. 
Thus, the galaxy velocity field of the infalling regions of clusters is expected to contain a significant radial 
infall component superposed to other orbits with larger angular momentum content.\\
The spherical infall model \citep{regos1989} describes the dynamical behavior of objects surrounding isotropic overdense regions with a collapsing velocity field whose 
amplitude depends on the distance (r) to the local overdensity \citep{diaferio1997}. 
\citet{peebles1976, peebles1980} derived a linear approximation to the velocity field induced by an isotropic mass overdensity ($\delta$) and the predicted infall velocity is
$V_{inf}^{lineal} = -1/3 H_0 \Omega_0^{0.6}$r $\delta$(r), where $\Omega_0$ is the density parameter and $H_0$ is the Hubble constant at present.\\ 
However, a pure spherical infall model cannot correctly predict the amplitude of the velocity field since in this model the amplitude of the velocity field depends on local conditions and not on the surrounding mass distribution.\\
Peculiar velocities arise through galaxy motions departing from a pure Hubble flow induced by the gravitational potential of mass overdensities distributed at large scales, so that global conditions are needed to fully describe the velocity field around a mass concentration.\\
Observationally, the effects of peculiar motions can be reliably inferred statistically via redshift-space distortions (hereafter RSD) studies \citep{croft1999, padilla2001, ceccarelli2006, paz2013, cai2016}. Besides, peculiar velocities can also be measured directly and used in other ways to extract useful information on the dynamics of galaxies and galaxy systems.\\
It should be recalled that peculiar velocity measurements directly trace
the mass distribution, avoiding the complications
of galaxy bias, present in RSD \citep{desjacques2010}. These measurements comprise a large range of scales and also may be used to add information on the nature of gravity by comparison to other modeling of galaxy motions.\\
For the mentioned reasons, the evolution of large structures can be constrained either by  
direct measurements of peculiar velocities or by the effects of redshift space distortions. Galaxy peculiar velocities generated by mass irregularities are superimposed to the cosmological 
expansion so that it can be easily inferred from its  redshift and a redshift independent distance estimation.\\
Along these lines, galaxy peculiar velocities are gaining interest as a promising cosmological probe
that provides new information on the dynamics of galaxies and systems at low redshift 
\citep{Johnson2014, huteler2017, dupuy2019, adams2020, kim2020}.
\citet{tonegawa2020} analyze redshift--space distortions in clustering measures to constrain cosmological parameters and
examine the satellite velocity bias between galaxies and dark matter inside haloes.
Recent works reconstruct peculiar velocities and the associated density fields using Cosmicflows-3 data. %
Their results highlight the ability of peculiar velocities to probe the mass distribution and reveal under/over--dense structures in the Local Universe \citep{Graziani:2019}.\\
Several authors have performed studies of the  
streaming motions at large and intermediate scales using peculiar velocities (in regions that extent up to $\approx$ 100 h$^{-1}$Mpc). In these studies, the bulk flow amplitude derived form the observational velocity field is often compared to the predictions of the standard $\Lambda$ CDM cosmological model
\citep{watkins2009, lavaux2010, feldman2010, colin2011, nusser2011, turnbull2012, ma2014}.

In this work we study the accretion of galaxies onto groups using the peculiar velocity field derived from redshift--independent distance measurements and analyze the infall amplitude dependence on mass group and surrounding galaxy density environment.
We also analyze the effects of anisotropies of the large scale galaxy distribution 
on the infall velocity fields onto groups.

This paper is organized as follows, 
in section 2 we introduce the data sets used,
in section 3 we describe the statistical method implemented to obtain the mean infall amplitude from observational peculiar velocities. In section 4 we display and analyze our results on infall onto groups using simulated and observational data respectively.
In section 5 we assess anisotropies on infall velocities. 
In section 6 we examine local effects vs large scale environment on infall. 
Finally, in section 7 we summarize and discuss our results.\\  
%
%
\section{Data}\label{sec:datos}
\subsection{Observational data}
The Cosmicflows--3 catalog (CF3) comprises almost 18.000 distances of galaxies and is the largest 
compilation of redshift independent extragalactic distances available \citep{tully2016}.
This catalog is based on the two previous versions, providing distances derived by the authors own observations as well as estimates extracted from the literature, homogenized to the same scale system.
Galaxy distances provided in the catalog are obtained from different methods, eg 
luminosity--linewidth  (Tully-Fisher)  relation,  the  Fundamental  Plane(FP),  surface-brightness  fluctuations,  Type  Ia  super-nova (SNIa) observations, etc. Nearby galaxy distances are accurate at the level of 5-10 $\%$. However, at larger distances the uncertainties raise up to 20-25 $\%$.\\
In this work, we use the online version of the CF3 consisting of data for groups of galaxies as well as individual galaxies with no association to groups. 
While the catalog include groups identified over the full redshift range of the the 2MRS survey \citep{hucra2012}, following the advice in \citet{tully2016}, we only considered the groups with velocities between 3,000 and 10,000 km$^{-1}$, since the properties of the nearest and farthest groups are uncertain. It is worth to mention that a group can have multiple contributions and therefore their distance uncertainty are reduced by averaging over all data source.

\begin{figure*}[h]
\centering
\includegraphics[width=.70\linewidth,height=0.35\textwidth]{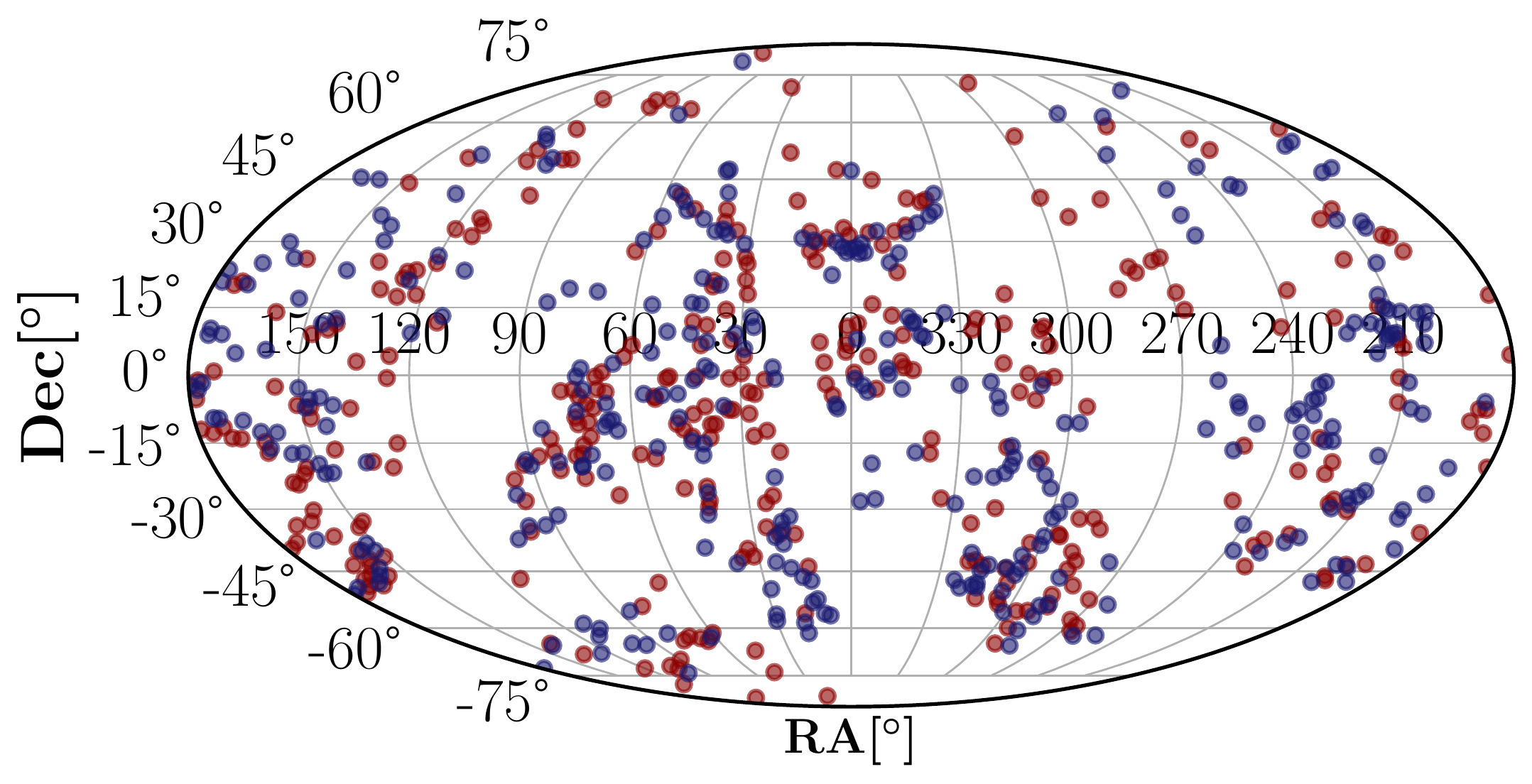}
 \caption{Sky distribution of galaxy groups in equatorial coordinates. Color codes correspond to high/low mass groups, M $>$ 3.4 x 10$^{13}$M$\odot$ (red circles), and M $<$ 3.4 x 10$^{13}$M$\odot$ (blue circles) respectively. }
\label{0_0a}%
\end{figure*}

The group catalog also includes masses estimated from the virial theorem and derived from the integrated $K_s$ band luminosity \citep{tully2015}.
It is well known that applying the virial theorem provides a poor estimate of group masses when these systems have a low number of members, while estimates of group masses through their total luminosities may provide more suitable proxies to the
actual masses (see \citet{eke2004}).
In addition, we highlight that only groups with at least 5 group members have mass estimates derived from the virial theorem.
For this reason, we use mass estimates derived from the integrated $K_s$ band luminosity throughout the paper.
In figure \ref{0_0a} we display the sky distribution of high and low mass groups, according to the median group mass value (red and blue circles for 
M $>$ 3.4 x 10$^{13}$M$\odot$ and M $<$ 3.4 x 10$^{13}$M$\odot$ respectively). As it can be noticed, both distributions trace large scale structures in a similar fashion.
Further details of group catalog can be found on \citet{tully2015, tully2016}.\\
The CF3 provides measurements of redshifts and distance moduli so that galaxy peculiar velocities, not explicitly included may be derived through: 
\begin{equation}\label{vpec}
   v_{pec} \approx (v_{mod} - H_{0}d) \diagup (1 - v_{mod}) 
\end{equation}
where v$_{mod}$ is the velocity with respect to the Cosmic Microwave
Background corrected for cosmological effects \citep{Tully:2013}.
The calibration of CF3 distances is set by the choice of the fiducial value of H$_0$ = 75 km s$^{-1}$. 
\citet{tully2016} shows that this value minimizes the monopole term with CF3 distances and results in a small global radial 
infall and outflow in the peculiar velocity field. For these reasons, we use this value in our analysis.\\
As it is well known, due to observational uncertainties that increase with distance, peculiar velocities can take unrealistic values.
For this reason, we have removed from our analysis galaxies with relative distance errors larger than 20 $\%$ and peculiar velocities larger than 1500km s$^{-1}$ ($|v_{pec}|$ > 1500kms$^{-1}$).\\
With these restrictions, the samples considered in this work comprise 2180 galaxies and 657 galaxy groups. This sample extracted from the CF3 is called hereafter CF3S.
In table \ref{table:samples} 
we summarize the galaxy and group samples 
used here and their main characteristics. 
The first two lines correspond to the CF3S galaxy and group samples previously described.
The following lines contain information on the group subsamples selected for the analyses 
carried out in this work and are introduced in the corresponding sections.  
\\

\subsection{Mock catalogs}
In order to compare model predictions to observational results we used mock catalogs based on galaxies extracted from the semi-analytical model presented by \citet{Henriques:2015}. 
This catalog include 2MASS photometric bands which are also in the group catalog supplement in CF3.
This semi--analytical model is an update of 
the Munich galaxy formation model consistent with the first--year Planck cosmology \citep{Planck:2011}, including a new treatment of baryonic processes to reproduce recent data on the abundance and passive fractions of galaxies from z= 3 down to z= 0.
\citet{Henriques:2015} use the Millennium Simulation \citep{Springel:2005},
which follows structure formation in a box of side 500 h$^{-1}$ Mpc comoving with a resolution limit of 8.6 $\times$ 10$^{8}\,h^{-1}\,$M$_{\odot}$.  
The cosmological parameters used in this semi-analytical model correspond to the  first--year Planck, $\sigma_{8}$=0.829, H$_{0}$=67.3 km\,s$^{-1}$ Mpc$^{-1}$,
$\Omega_{\Lambda}$ = 0.685, $\Omega_{m}$ = 0.315, $\Omega_{b}$=0.0487.\\
We have constructed 25 mock catalogs in order to compare 
model predictions to observational results and examine possible biases generated by systematic errors in the observational data. Each of these mock catalogs have a
Hubble parameter H$_{0}$=75 km\,s$^{-1}$ consistent with the Cosmicflows--3 catalog and are used to estimate the effect of cosmic variance on our results.
The set of mocks mimic as much as possible the characteristics of the observational catalogs.
In order to reproduce the statistical properties of the observations we place an observer inside the simulation box, and consider a volume representative of the observational volume.
The same angular mask as the observational catalog is applied, excluding an area similar to that produced by the extinction of our galaxy's disk (Zone of Avoidance).
The galaxies are selected at random in such a way to obtain comparable number density of galaxies with a similar redshift distribution.
In addition, the mock catalogs are also designed to reproduce
the current observational measured distances and their estimated uncertainties.\\

\begin{figure*}[h] 
 \hfill
 \subfigure{%
   \includegraphics[width=.40\linewidth,height=0.35\textwidth]{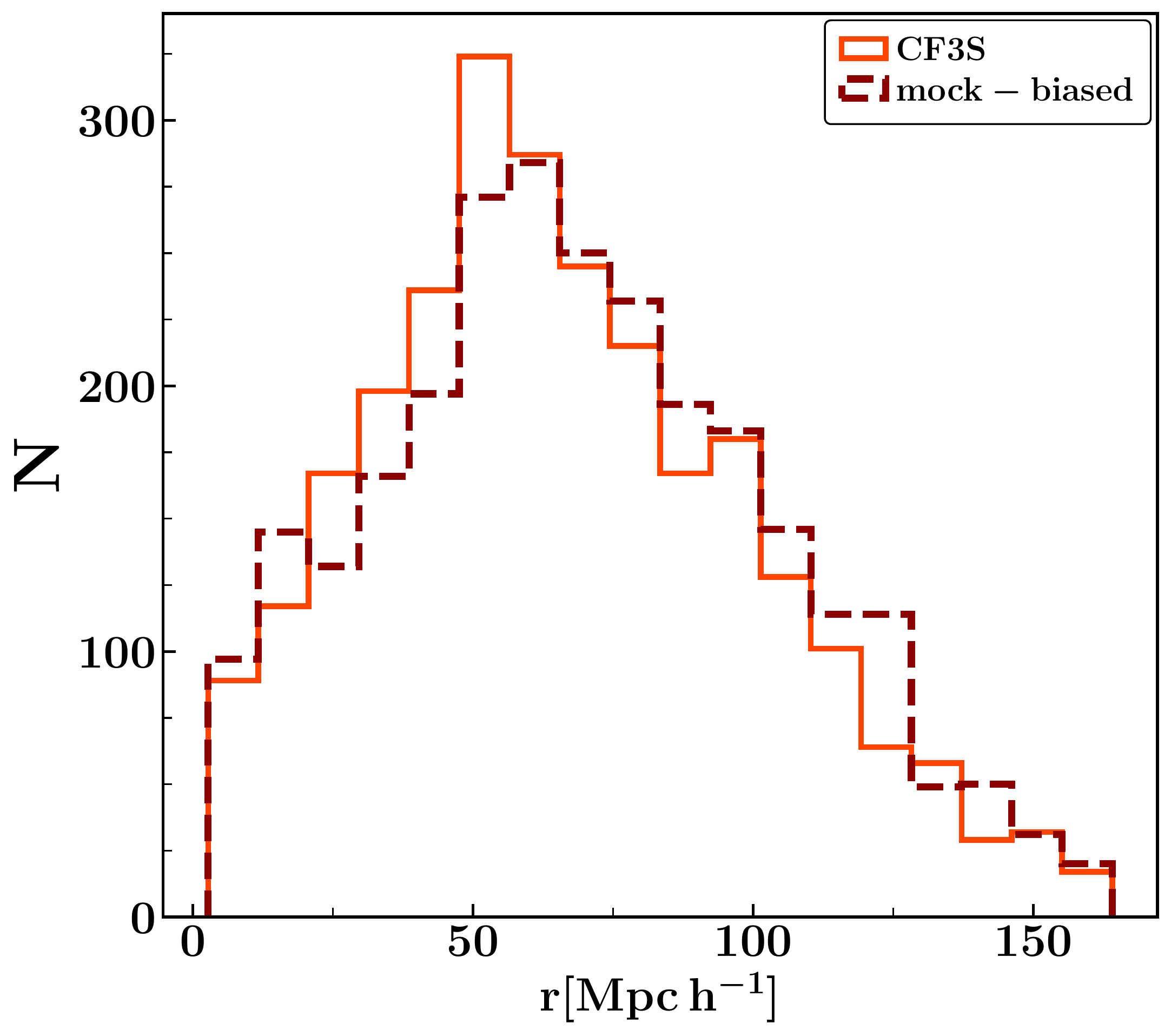}}~\hfill~
 \subfigure{%
 \includegraphics[width=0.40\textwidth,height=0.35\textwidth]{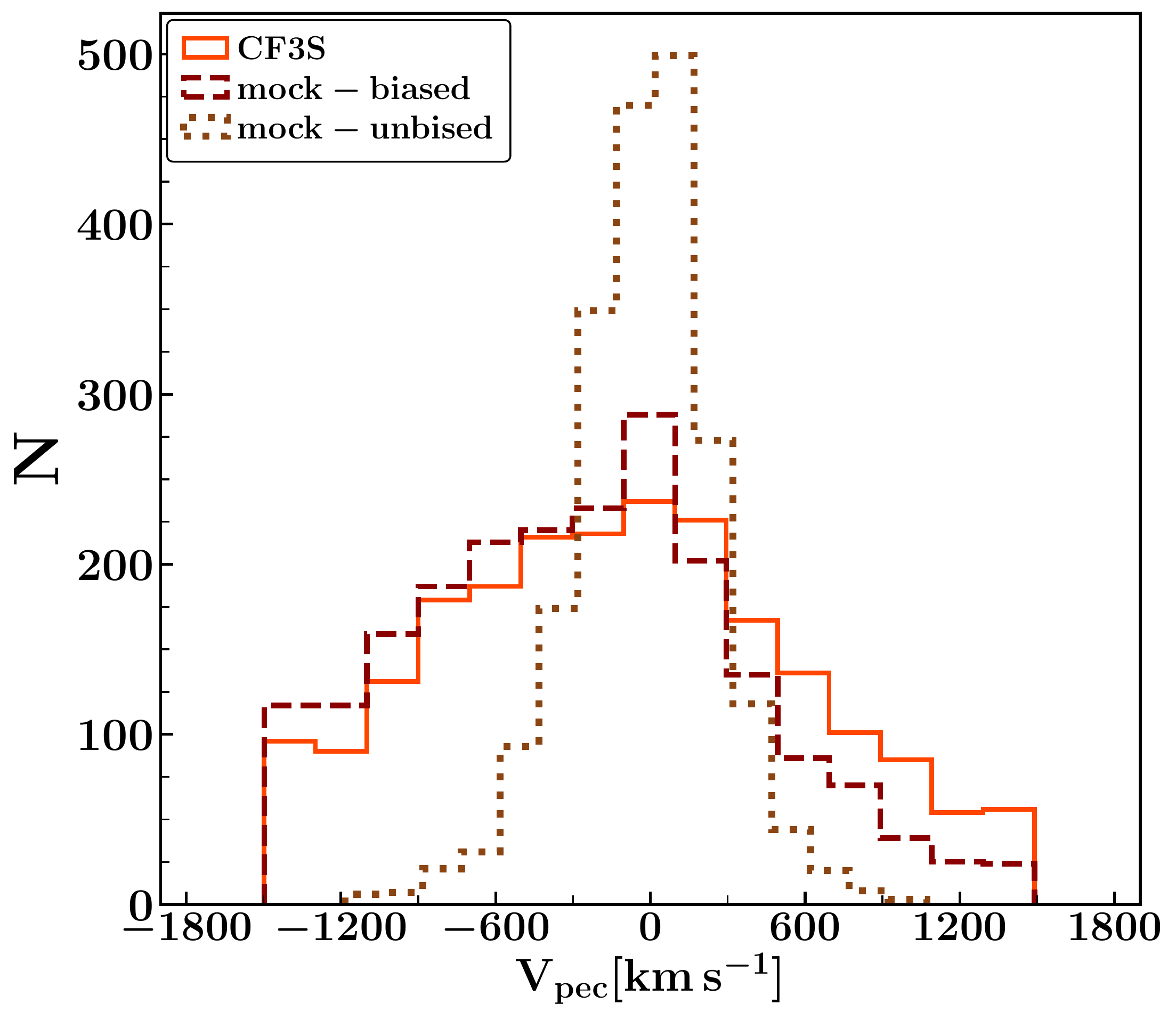}}~\hfill~
 \caption{Left panel: Distributions of distances in the Cf3S (solid orange line) and in a typical mock catalog with distance errors included (mock--biased, dashed brown line). 
Right panel: Distribution of radial peculiar velocities in the CF3S (solid line histogram) and in the 
mock biased catalogs (dashed histogram). A large tail towards negative values 
is observed in both distributions.
The distribution associated to the mock catalog without errors (mock--unbiased, dotted brown line histogram) has a nearly Gaussian radial peculiar velocity distribution.  
}
\label{0_0b}%
\end{figure*}

We use haloes identified with a ‘friends of friends’ (FOF) algorithm with standard parameters \citep{Henriques:2015}. We 
select those FoF halos imposing a similar mass distribution as the observations.
These restrictions provide suitable samples to test the dynamics of infalling semianalytic galaxies onto these mass inhomogeneities, resembling the infall of galaxies onto groups in the real universe.\\
Large uncertainties of distance measurements strongly affect the derived radial peculiar velocities and uncertainties can be as large as the peculiar velocities. 
In addition, distance estimates are subject to 
systematic biases, as homogeneous and inhomogeneous Malmquist biases \citep{Strauss:1995,Dekel:1994}, that can generate spurious artifacts in the inferred velocity field.\\
We have analyzed and quantified the impact of uncertainties of distance measures on 
the infall mean velocity determination by assigning errors to the mock catalogs similar to those affecting the observational data\\ In general, the different distance indicators have a fractional distance  uncertainty. Primary distances estimators such as 
the tip of the giant branch, cepheids, or surface brightness fluctuations give distances 
with relative errors up to 5\% $-$ 10\% whereas secondary indicators,
reach uncertainty estimates up to 25\%.\\      
As it is well known, uncertainties in redshift independent galaxy distance estimates are consistent with a Gaussian distribution in distance moduli.
As a consequence of this behavior, uncertainties of both, distance and peculiar velocity 
are expected to have a log--normal distribution.
In order to asses the effects of these distance uncertainties on our results, 
we have assumed a gaussian distribution of errors in distance moduli, with mean and dispersion as derived from the CF3, ie. consistent with a mean relative distance uncertainty of 17\%.  
These uncertainties are used to modify the distances of each galaxy in the mock catalog d$_{n}$ = d$\times$10$^{f_d}$, where $f_d$ is taken from a Gaussian distribution.
These mock catalogs are called hereafter mock--biased catalogs, while mock catalogs with no errors in galaxy distances, mock--unbiased catalogs.\\
The left panel of figure \ref{0_0b} displays both distance distribution for the CF3S 
and for a typical random mock with modified distances (mock--biased). 
As it can be seen in this figure both distribution are similar, providing confidence on our assumptions to assign distance uncertainties in simulated data.
The radial peculiar velocities computed from equation \ref{vpec} are shown in the right panel. Dashed lines correspond to the CF3S and dotted--dashed line to the mock--biased. 
In order to highlight how distance errors introduce biases in radial peculiar velocities
we show in the same panel the distribution of the velocities for the same
mock with error free distances (mock--unbiased). The distributions related with CF3S and mock-biased reveal a skewness towards
negative peculiar velocities and both are flatter than the mock--unbiased.
This effect is a consequence of the asymmetry in the distribution of fractional 
errors on distance, skewing peculiar velocity measurements to negative values.
Regarding comparison between distances and peculiar velocities, there is a qualitatively 
good agreement between CF3S and mock--biased catalog. The galaxies from the mock--biased catalog
are distributed consistently with observations.
                       
\begin{table*}[h]
  \begin{center}
   \caption{Summary of the main characteristics of groups and galaxy samples.}
    \begin{tabular}{cccccc}
      \hline
      \hline
      \noalign{\vglue 0.2em}
      sample & object & selection criteria & mass range [M$_{\odot}$] & vel range [km s$^{-1}$] & Number\\
      \noalign{\vglue 0.2em}
      \hline
      \noalign{\vglue 0.2em}
     CF3S & galaxies & & & 100--15000 & 2180  \\
      all & groups &  & [2$\times$10$^{12}$ 
      - 5.5$\times$10$^{14}$]
      &3000--10000 & 657\\
      HM & groups & group mass & $>$3.4 10$^{13}$ & 3000--10000 & 328\\
      LM & groups & group mass & $<$3.4 10$^{13}$ & 3000--10000 & 329 \\ 
      \noalign{\vglue 0.2em}
      \hline
      \noalign{\vglue 0.2em}
       & &  &  direction & & \\
      \noalign{\vglue 0.2em}
      \hline
      \noalign{\vglue 0.2em}
      $\parallel$ overdensities & groups & large scale anisotropies & $\parallel$ LOS & 3000--10000 & 296 \\
      $\perp$ overdensities & groups & large scale anisotropies & $\perp$ LOS& 3000--10000 & 361\\
      \noalign{\vglue 0.2em}
      \hline
      \noalign{\vglue 0.2em}
       & &  &overdensity range & & \\
      \noalign{\vglue 0.2em}
      \hline
      \noalign{\vglue 0.2em}
      high density & groups & overall galaxy density & $>$4 & 3000--10000 & 150 \\
      intermediate density & groups & overall galaxy density & 2--4 & 3000--10000 & 225 \\
      low density & groups & overall galaxy density & 0--2 & 3000--10000 & 271 \\
      \noalign{\vglue 0.2em}
      \hline
      \hline
    \end{tabular}
    \label{table:samples}
     \label{table:samples}

  \end{center}
\end{table*}

%
\section{A statistical approach to derive the mean infall component of galaxy peculiar velocities onto groups}\label{metodo}
From the dynamical point of view, galaxy groups can be idealized as spherical, isolated mass overdensities. Under this simplified approximation, the expected 
peculiar velocity field of galaxies in groups and in their surroundings results from the sum of a spherical infall component plus a random velocity dispersion induced by the virialization of the central region.   
This picture is consistent with linear theory where convergent velocity fields may be directly associated to spherical isolated mass overdensities in a cosmological substratum.

Given the limitations of the observations where only the line--of--sight projection (LOS) of the peculiar velocity of galaxies can be estimated, the derivation of the observed velocity field associated to the infall onto groups depends on the position of the galaxy relative to both, the group center and the observer (see equation \ref{coseno}).
Then, for galaxies at relative distance $r$  to the group, the infall amplitude ($V_{inf}(r)$) can be derived from the LOS peculiar velocity ($V_{pr}$) by the lineal relation shown in equation \ref{coseno},
\begin{equation}\label{coseno}
V_{pr}(r, \theta) = V_{inf}(r)\cos(\theta) 
\end{equation}
where $\theta$ is the angular separation between the galaxy and the observer as seen from the group center.\\
In order to derive $V_{inf}(r)$, we consider galaxies located at different spherical concentric shells of radius $r$ around the groups.
For each shell we examine the dependence of the peculiar velocity on cos($\theta$).
Then, for different cos($\theta$) bins we calculate the average galaxy peculiar velocity
$\langle V_{pr}\rangle$, removing iteratively the galaxies those peculiar velocity 
lie at more than 1.5$\sigma_{V_{pr}}$ from the mean derived for each bin with 
the data used in the previous iteration. 
The mean infall amplitude V$_{inf}(r)$  can be simply derived 
by applying a least--squares linear fit to the (cos($\theta$),$\langle V_{pr}\rangle$) vs group--centric radius r.\\
Note that in the reference system adopted, positive velocities correspond to infall motions 
while outflow velocities have negative values in our convention.
%
%
\begin{figure*}[h] 
 \hfill
 \subfigure{%
   \includegraphics[width=.50\linewidth,height=0.40\textwidth]{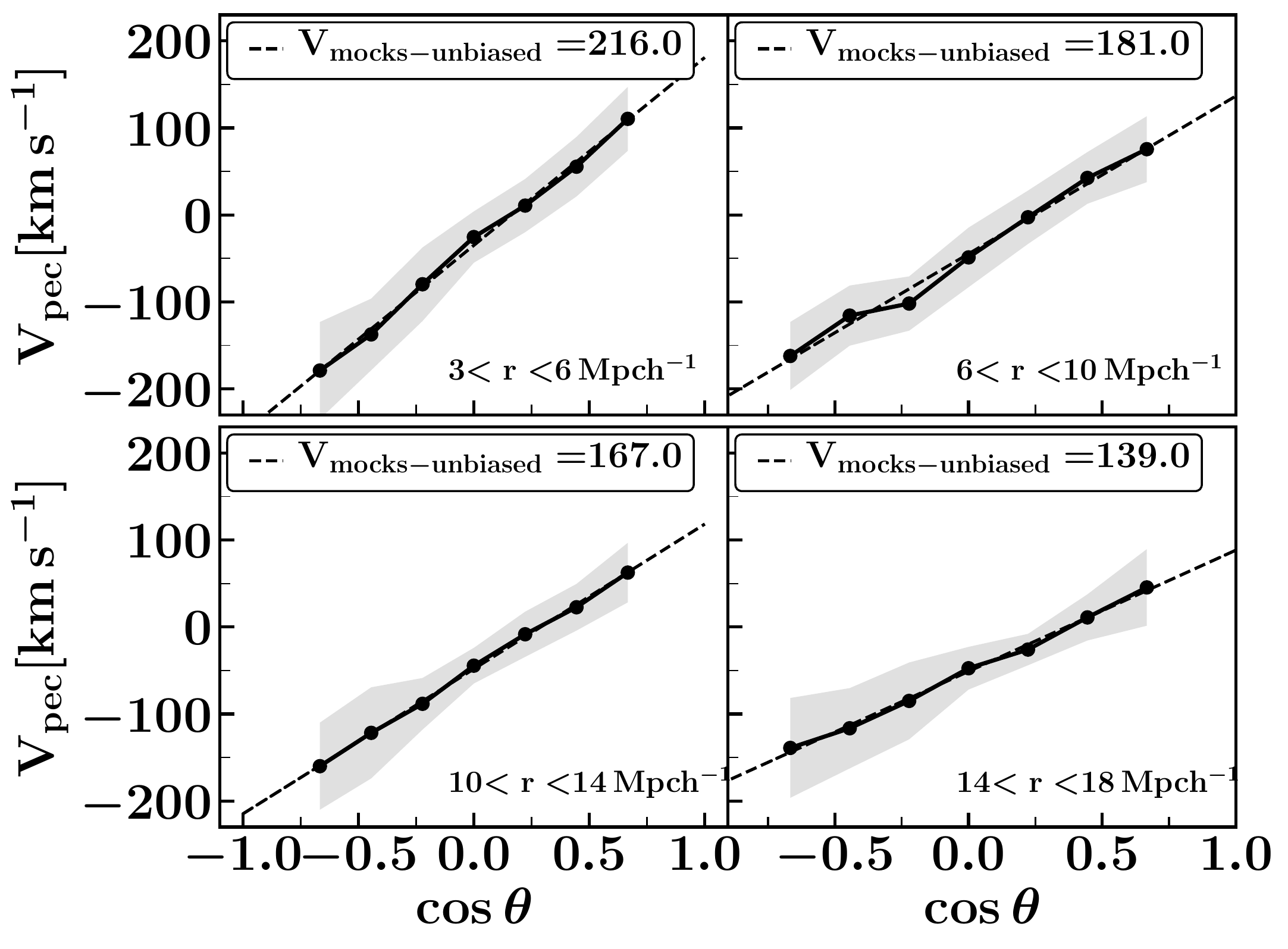}}~\hfill~
 \subfigure{%
 \includegraphics[width=0.5\textwidth,height=0.40\textwidth]{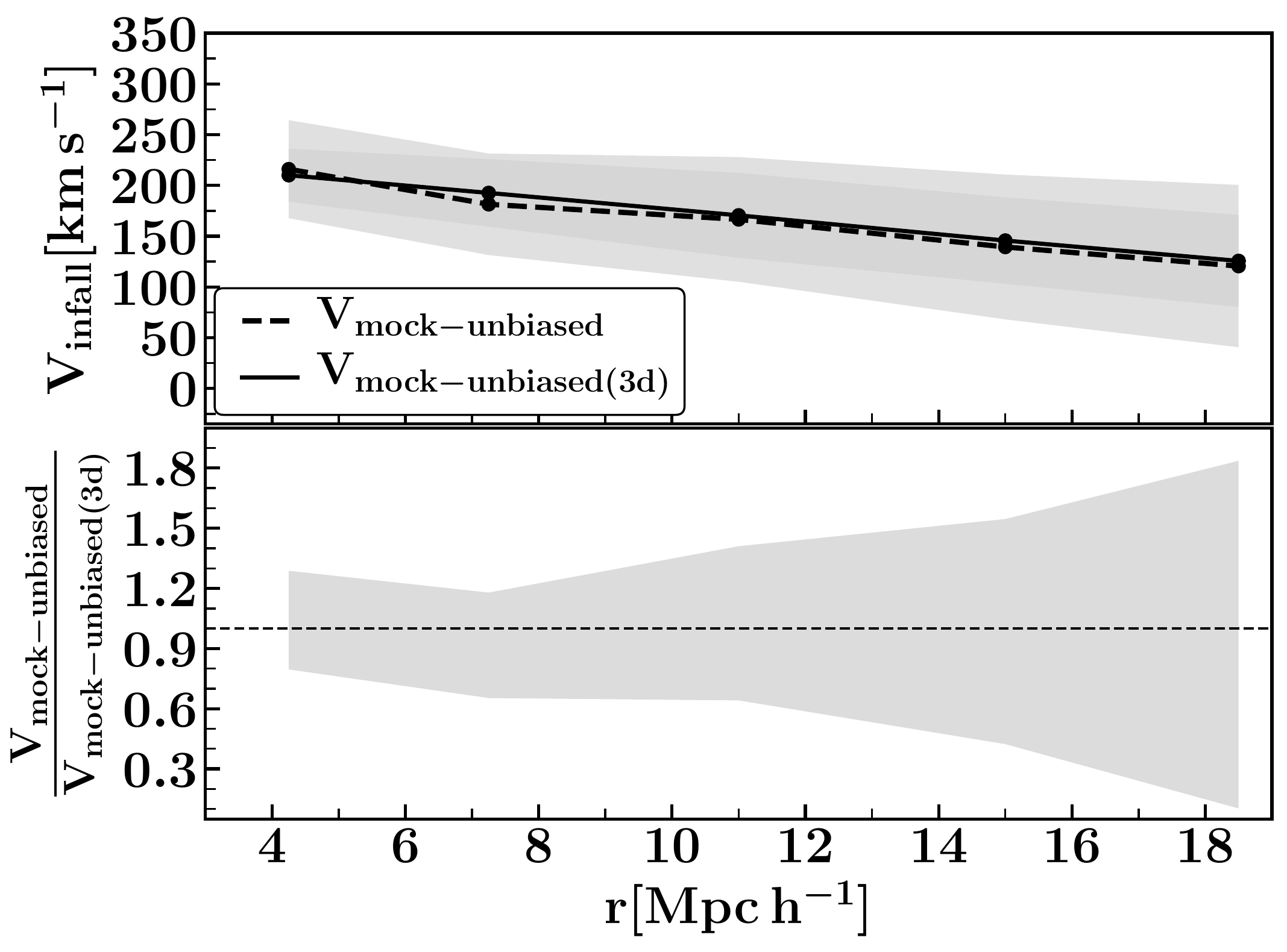}}~\hfill~
 \caption{
 Left figure:
 Mean peculiar velocity as a function of the angle the galaxy--group and group--observer directions ($\Theta)$ in the mock--unbiased catalogs. Each panel correspond to a different galaxy--group distance, as is indicated in the figure. The dashed line indicates the best linear fitting, the slope is associated to the infall amplitude (V{\scriptsize mock--unbiased} ) and it is 
 specified in each panel. The gray region correspond to the peculiar velocity dispersion. 
 Right figure, upper panel: Mean infall velocity as a function of distance to the group center in the 
mock--unbiased catalogs. 
Solid line indicates the mean velocity directly obtained from the 3-dimensional peculiar velocity (V{\scriptsize mock--unbiased(3d)}).   
Dashed line corresponds to the infall amplitude (V{\scriptsize mock--unbiased}) obtained by applying our statistical method at LOS peculiar velocities and 
coincides with the slopes of the lines in the left figure .
The displayed velocities are averaged over 25 mock--unbiased catalogs and the shadow 
region shows their dispersion.
Right figure, lower panel: Dispersion of the ratio between the mean infall velocities V{\scriptsize mock--unbiased(3d)} 
 and the mean infall V{\scriptsize mock--unbiased} velocities for each mock--unbiased}.
\label{1_0}%
\end{figure*}
\subsection{Recovering the mean infall amplitude in simulated data}\label{sec:resultsim}
The comparison between the mean infall amplitudes as a function of r obtained by averaging the peculiar velocities
and those derived form equation \ref{coseno} allows for a reliability test of the our methods.
For this aim, we take advantage of the information provided by synthetic catalogs, in particular the three-dimensional and the LOS peculiar velocities.\\
We explore the dependence of the peculiar velocity on cos($\theta$) in the mock--unbiased catalogs
and apply a least-squares linear fit to (cos($\theta$),$\langle V_{pr}\rangle$).
We show some examples of this relation in the four panels at the left of figure \ref{1_0}, where each panel corresponds 
to different group--centric distance range, as indicated in the figure, where the slopes of the
linear fits are displayed in the small upper boxes in each panel.
These slopes correspond to the derived mean infall amplitudes and are used
to generate the infall curve shown in the right panel of this figure.\\
We derive the mean infall amplitude V$_{inf}(r)$ as a function of distance to the group for the 
25 mock catalogs and in the upper right panel of the figure \ref{1_0} we show the results averaged 
over the 25 mock--unbiased catalogs.  
The solid line denotes the mean infall motion derived from 
the projection of the three--dimensional velocity vector  
(V{\footnotesize mock--unbiased(3d)}) 
along the group--centric direction, whereas the dashed  
line corresponds to the mean infall velocity 
(V{\footnotesize mock--unbiased})
inferred by
the procedure described in section \ref{metodo} and shown in the left panels.\\
Uncertainties are derived from the scatter of measurements obtained from the 25 mock--unbiased catalogs, consistent with a suitable measure of cosmic variance.\\
As it can be seen in upper right panel of figure \ref{1_0}, the mean infall velocity estimate via the three--dimensional vector 
(solid line) and those inferred from LOS peculiar velocities (dashed line) 
are in excellent agreement, indicating that the proposed statistical method 
allows to derive the mean infall velocities from peculiar velocities.\\
We notice a convergent velocity field onto groups which can be 
clearly distinguished up to group--centric distances of 
18 Mpc h$^{-1}$, showing large scale flows of  galaxies directed onto groups.\\
Closer to group centres, the mean infall velocity reaches a maximum of approximately 200 km s$^{-1}$,
where the effects of the potential well associated to the groups dominates the local radial infall. At larger distances the velocity field is no longer dominated by the group mass concentration but rather obeys the surrounding large scale structure, implying an average decline of the mean infall velocity.\\
The shaded region in the lower right panel of figure \ref{1_0} shows the dispersion of the ratio between the actual mean group--centric infall velocity and the mean infall velocities derived from peculiar velocities for each mock--unbiased. As we can see there is a suitable agreement between the two measures which provides confidence in our method in deriving reliable infall amplitude determinations.
%
%
\begin{figure*}[h]
\centering
\includegraphics[width=0.9\textwidth,height=0.65\textwidth]{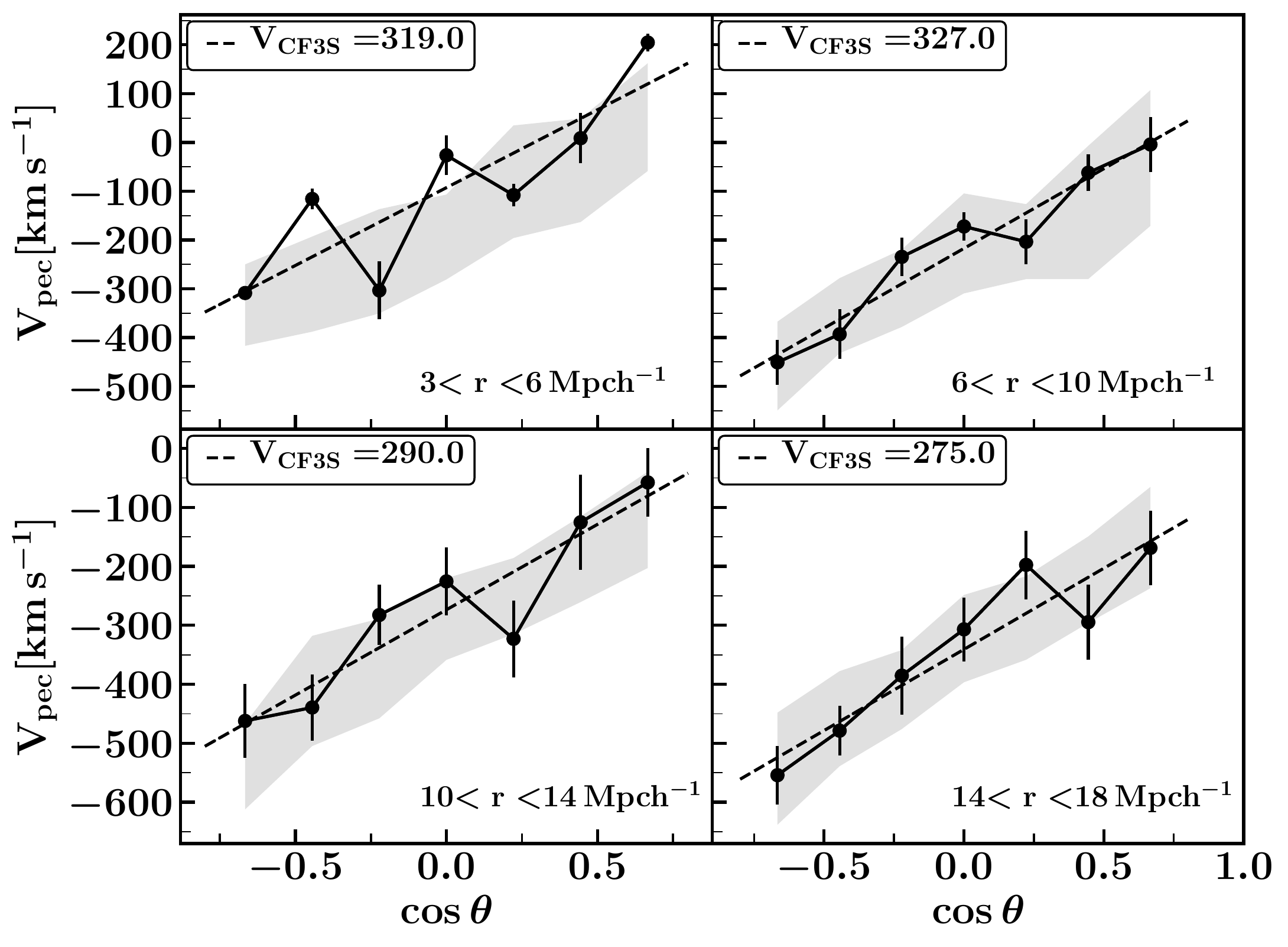}
\caption{
Mean peculiar velocity as a function of the angle between the directions galaxy-group and group-observer ($\theta$) in the CF3S (points). Each panel correspond to a different galaxy-group distance, as is indicated in the figure. The dashed line indicates the best linear fitting, the slope is associated to the \textit{infall} amplitude (V{\scriptsize CF3S}) and it is specified in each panel. The error--bars correspond to the peculiar velocity dispersion.
The gray shaded areas encloses the dispersion of mean peculiar velocity from 25 mock--biased catalogs. Notice that the range of y-axis in upper and lower panels is different.}
\label{1_4}
\end{figure*}
%
%
\section{Observed velocity field around groups} 
Once we have tested the reliability of our methods to assess the effects of LOS projection of peculiar velocity in the mock--unbiased catalogs, we study here the streaming infall flow around groups. For this aim we use 
observed peculiar velocity data applying our analysis to the samples of galaxies and groups described in section \ref{sec:datos}.\\
We derive the mean streaming velocity toward groups in the observational data by applying the same procedures used in the mock--unbiased catalogs (section \ref{sec:resultsim}). 
In figure \ref{1_4} we show the mean projected peculiar velocity as a function of cos($\theta$) bins for the samples of galaxies and groups taken from the CF3S (points) and their best linear fitting (dashed lines).  
The different panels correspond to spherical shells at different distances (r) 
to the groups, as indicated in the figure.   
In this figure, uncertainties are derived from the scatter of measurements obtained from the CF3S (error--bars), and gray shadow regions correspond to the dispersion of results obtained in 25 mock--biased catalogs. We acknowledge the very good agreement between the results of the mock--biased catalogs and the observations, totally consistent within uncertainties.\\
%
%
\begin{figure}
\centering
\includegraphics[width=1.0\linewidth,height=0.8\hsize]{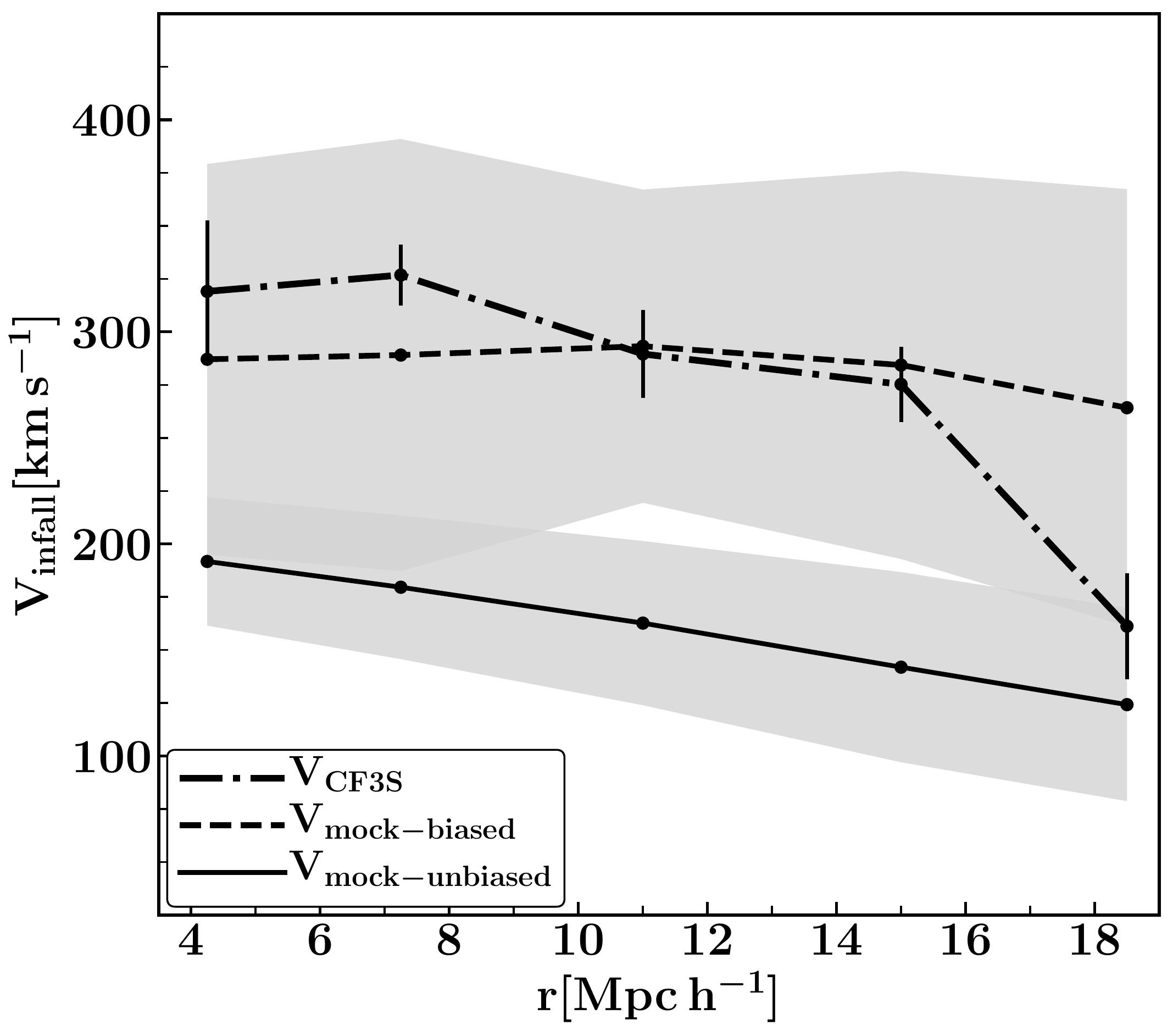}
\caption{Mean infall velocity as a function of distance to the group center for biased samples.    
Dashed line indicates the mean velocity obtained from the mock with errors included (mock-biased).   
and the dot--dashed line corresponds to Cosmicflows-3 sample (CF3S). For comparison we include the mean infall velocity 
for mock-unbiased (solid line).
Error bars correspond to the uncertainty in the mean infall velocities from CF3S. 
The shaded region indicates the variance obtain from 25 mock without (with) errors included catalogs realizations.
}
\label{1_2}
\end{figure}
%
As it can be seen, we obtain positive slopes consistent with a neat infall (see the small 
panels in the right of the figure) for all the distance ranges analyzed. 
Thus, our results show a clear signal for infall 
onto groups. Moreover, by inspection to this figure it can be noticed that the 
dispersion around the mean infall velocity is larger in the inner group--centric distance bin (upper left panel of the figure \ref{1_4}) possibly due to the contribution of motions from the virialized and pre--virialized regions \citep{diaferio1997,diaferio1999}. %
Beyond this scale the mean projected peculiar 
velocity become more stable and errors get smaller, showing the expected smoother flow from large distances.
At larger scales the influence of surrounding structures starts to affect the infall pattern and the systematic radially inward velocities
decreases. This behavior is expected at large scales where the infall signal becomes negligible due to the presence of the gravitational pull of neighboring groups, clusters and filaments 
(notice the different peculiar velocity ranges of the upper and lower panels).\\
This behavior could be a consequence of biases that affect 
CF3S distances resulting in spurious velocity flows \citep{tully2016}. Even more, the radial peculiar velocity distribution has a negative skewness, as shown in figure \ref{0_0b}, and this asymmetric bias could contribute 
to obtain large negative peculiar velocities.
It is worth to notice that mock--biased catalogs present the same behavior (gray shaded areas)
while mock--unbiased do not show this trend.\\
%
%
\begin{figure}
\centering
\includegraphics[width=1.0\linewidth,height=0.8\hsize]{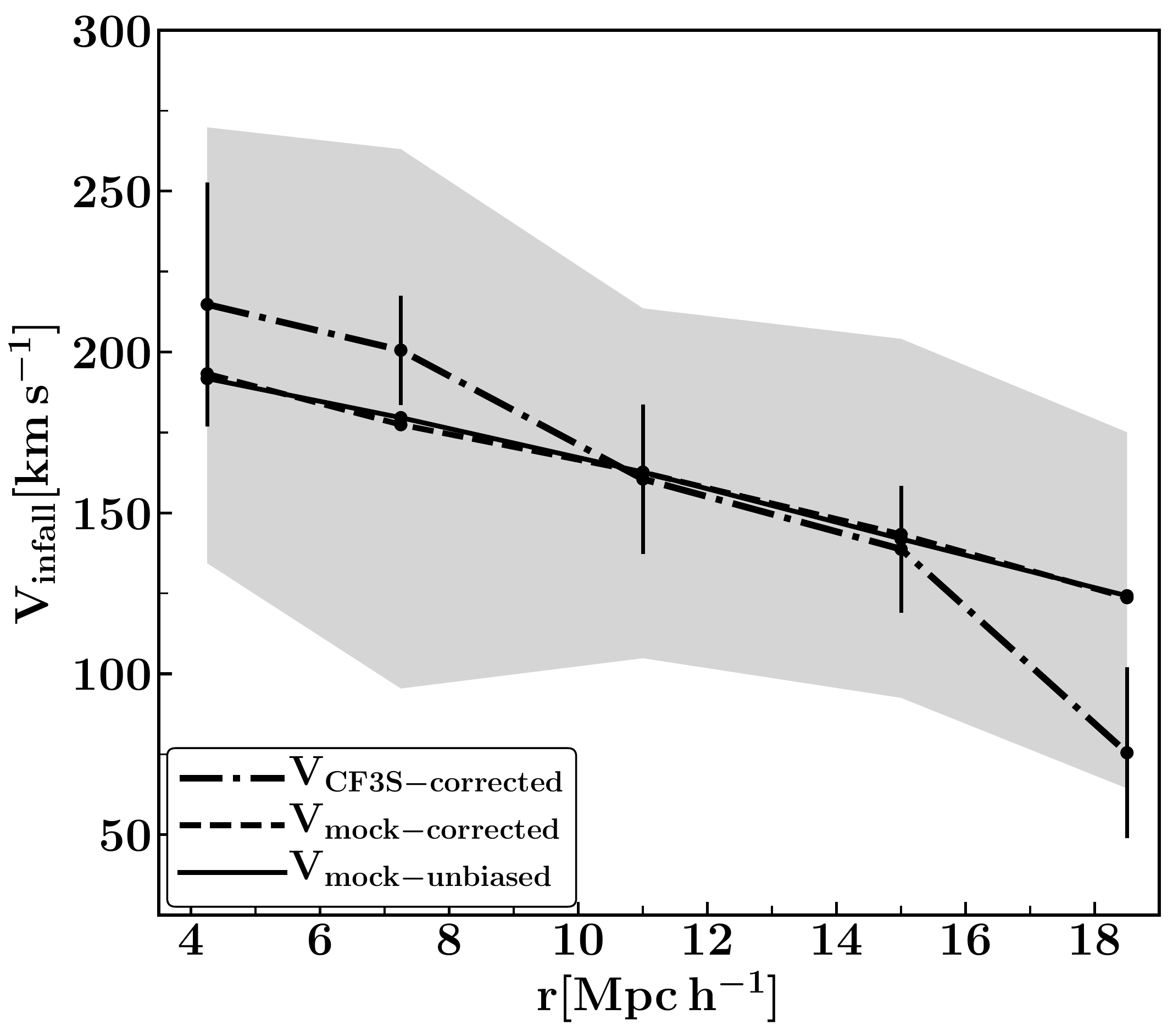}
\caption{Mean infall velocity as a function of distance to the group center for corrected samples. 
The dashed lines indicate the mean velocity obtained from the corrected mock catalogs, and the dot--dashed line 
corresponds to Cosmicflows--3 sample corrected (CF3S--corrected). 
The solid lines corresponds to the mean infall velocity for the mock--unbiased catalogs shown in solid lines of figure \ref{1_2}.
Error bars correspond to the uncertainty in the V{\scriptsize  CF3S--\footnotesize corrected}.
The shaded regions indicate the variance obtain from 25 mock--corrected catalogs.
}
\label{1_5}
\end{figure}
In figure \ref{1_2}, we show the resulting mean velocity infall as a function of group--centric distance, V{\footnotesize infall(r)} for CF3S, (V{\scriptsize CF3S}, dot--dashed lines), as 
derived from the linear fit applied to each panel of figure \ref{1_4}. We also show in this figure the results from mock LOS peculiar velocities with  (V{\footnotesize mock--biased}) and without (V{\footnotesize mock--unbiased}) errors included (dashed and solid lines respectively).\\
It can be seen that when distance uncertainties are included,  the mean infall velocity  corresponding  to  mock--biased  catalogs,  
agree well with those derived from CF3S data (dot--dashed and dashed lines respectively). 
By inspection to figure \ref{1_2} it can be seen that V{\scriptsize CF3S} infall flow (dashed line) reaches a maximum of approximately 300 kms$^{-1}$, which is 50\% larger when compared to 
the infall flow from mock--unbiased catalogs  (solid line).\\
As it is evident from this analysis, a proper address of distance uncertainties is crucial in deriving reliable peculiar velocity fields around groups.

\subsection{Correcting for distance uncertainties effect on infall determination}

The resulting average group--centric infall 
velocities recovered from the mock catalogs are used to calibrate and correct the observational results.\\
In order to correct for the effect of distance uncertainties in the inferred infall velocities, 
we compare the results from the mock catalogs with and without the inclusion of errors (mock--biased and mock--unbiased respectively).
We calculate the ratio f $ = $ V{\footnotesize mock--biased} $\diagup$ V{\footnotesize mock--unbiased}, where V{\footnotesize mock--biased} is 
the infall amplitude of the mock catalogs with distance errors (see dashed line in 
figure \ref{1_2}) and V{\footnotesize mock--unbiased} is the infall amplitude from the mock catalogs without 
distance uncertainties (see solid lines in figure \ref{1_2}).\\ 
A fitting function of the form \textit{f} $=$ \textit{a}$r^2$ + \textit{b}$r$ + \textit{c} is sufficient 
to provide a good description of the ratio of observed to actual velocities. 
We find that the parameters \textit{a} = 1.2, \textit{b} = 0.05, y \textit{c} = - 0.0002 provide a good fit to the measured values of f for this group sample.
The fitting parameters obtained for different subsamples provides the factors ($f$)
used to correct the infall velocities measured (V{\scriptsize CF3S}) in the corresponding CF3S.
These observational flows corrected hereafter are called V{\scriptsize  CF3S--\footnotesize corrected}, in a similar way flows corrected on 
mock-biased catalogs are called V{\footnotesize mock--corrected}.
The corrected infall pattern derived from the CF3S is shown as dot--dashed lines in
figure \ref{1_5}, the dashed line corresponds to the average infall flow in 25 mock--corrected catalogs, and the 
solid line corresponds to the actual mean infall flow directly obtained from peculiar velocities in the mock--unbiased catalogs. 
The gray region enclose the 5 and 95 percentiles of the infall distribution on the 25 mock-corrected 
catalogs which can be taken as a measure of cosmic variance.\\
As can be seen in figure \ref{1_5}, there is a clear evidence of infall motions up to 
16 Mpc h$^{-1}$.
We also  notice that V{\scriptsize  CF3S--\footnotesize corrected} and V{\footnotesize mock--corrected} reaches 200 kms$^{-1}$ at 
regions close to the groups ($\approx$ 4 Mpc h$^{-1}$).
The resulting mean velocities obtained here are consistent with those found by 
\citet{ceccarelli2005}. 
For larger distances to the groups the infall amplitude decreases down to about 120 kms$^{-1}$ and infall uncertainties increases. \\
Here and throughout, we used the Kolmogorov--Smirnov (KS) test to find the statistical significance of the differences between infall velocity amplitudes for the different samples.  We consider that the differences between distributions are highly significant if the p--value es p<0.01.\footnote{When we quote the statistical significance of a difference between two distributions (x$_i$ and y$_i$), we compute 
 P($\Delta \chi^2$, N$_{dof}$) where $\Delta \chi^2 = \sum_i [d_i/\sigma(d_i)]^2$ with d$_i$ = (x$_i$-y$_i$) for each bin and $\sigma(d_i) = \sqrt{\sigma(x_i)^2+\sigma(y_i)^2}$. Notice that P(x, N) is a $\chi^2$ distribution with N degrees of freedom.}
The KS test confirms  that the slope value of V{\scriptsize  CF3S--\footnotesize corrected} profile in figure \ref{1_5} is significant at the 3 $\sigma$ level.\\
These results are reasonable considering that close to the groups, galaxy peculiar velocities are dominated 
by the group potential well whereas at larger distances the surrounding 
large scale structure strongly affects the velocity field.\\
Taking into account our cosmic variance determinations, the mean infall 
velocities V{\scriptsize  CF3S--\footnotesize corrected} are consistent to those V{\footnotesize mock--corrected} from mock--biased catalogs.
\subsection{Dependencies on group properties}
In this subsection, we analyze the characteristics of the peculiar velocity field around groups and its relation to group mass.\\
We subdivide the group sample according to mass estimated from the integrated $K_s$ band luminosity (hereafter mass, M) into two group subsamples: groups with M $>$ 3.4 x 10$^{13}$M$\odot$ and M $<$ 3.4 x 10$^{13}$M$\odot$
dubbed high Mass (HM) and low Mass groups (LM) respectively. 
This limit was chosen in such way that each subsample contains a similar numbers of groups, as shown in table \ref{table:samples}.\\
Similarly, we analyze the 25 mock catalogs and estimate the mean infall amplitude for galaxies in the mock--unbiased and mock--biased catalogs. 
This allows to infer the correction factors ($f$) for each sub-sample and use them to derive the true amplitudes of the infall velocity field in the observations.
Here and throughout, observed mean infall velocities always refer to those corrected by the factor f in CF3S and the cosmic variance is obtained from the 25 mock--corrected catalogs.
%
%
\begin{figure}
\centering
\includegraphics[width=1.0\linewidth,height=0.8\hsize]{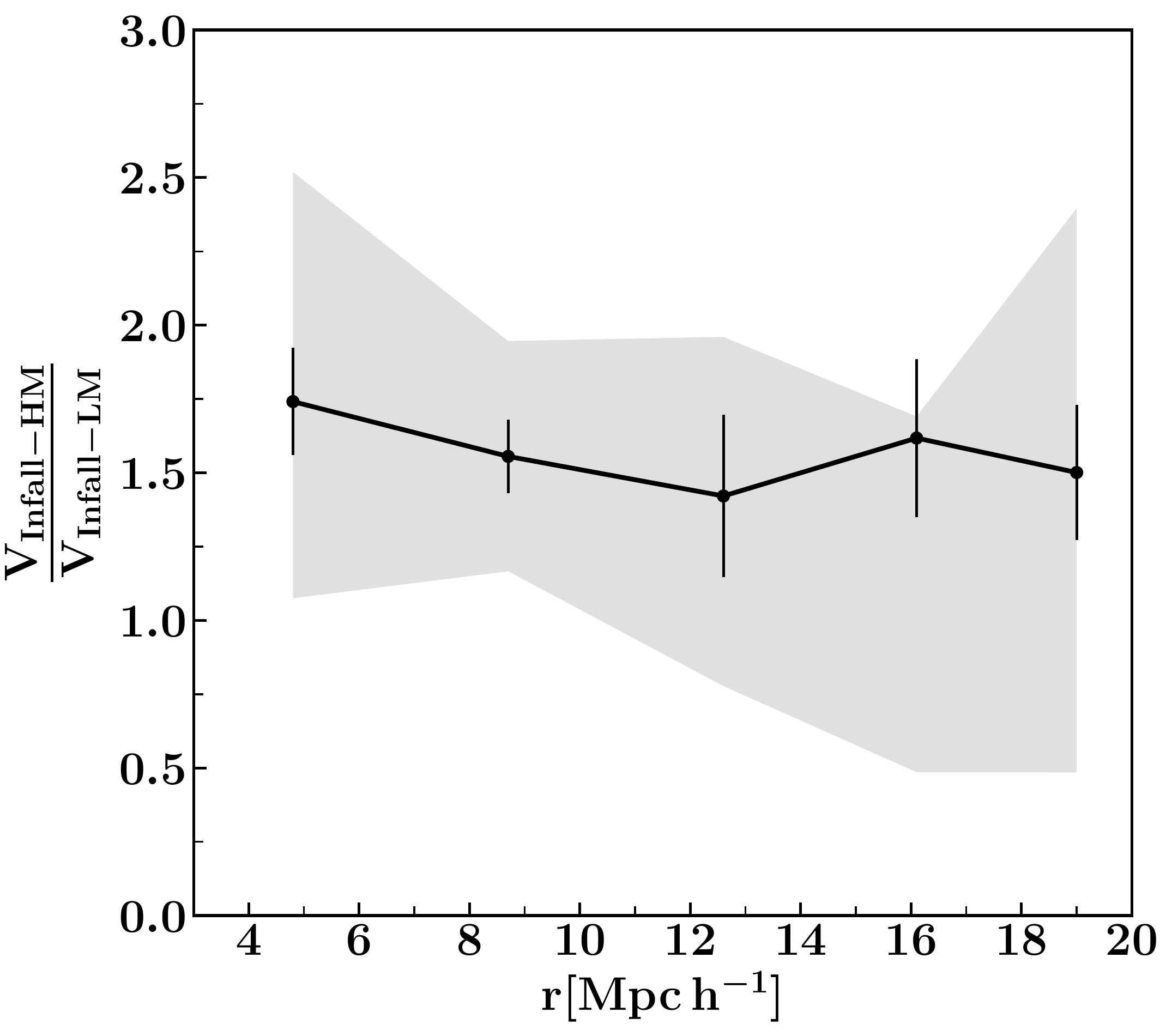}
\caption{The solid line shows the ratio of the mean infall velocity of high to low mass groups as a function of distance to the group centers in observational data.
Error-bars indicate the uncertainty in the mean infall velocity and the gray shadow region shows the dispersion of 25 mock catalogs.}
\label{1_6}
\end{figure}

The resulting relative amplitude of the radially inwards streaming motions are shown in figure \ref{1_6}. We show 
the ratio ($\frac{"HM"}{"LM"}$) between the infall amplitude of the two samples. 
As it can be seen in this figure, there is a clear difference in
the mean infall amplitude onto group samples of high and low mass,  
the infall velocity amplitude associated to the high mass groups are systematically larger than that 
corresponding to low mass groups in the range of group centric distances explored.
This behavior is stronger close to the center of the groups, where  
galaxies around high mass groups exhibit noticeably higher infall than galaxies around low mass groups. 
The KS test shows that the differences between infall velocity amplitudes are statistically significant (p<0.01), 
consistent with a statistical confidence at the 3 $\sigma$ level.
The gray shaded area in this figure corresponds to variance cosmic taken from the mock catalogs while error--bars indicate 
the relative uncertainty between both samples as derived from the linear fitting of peculiar velocities and cos($\theta$).\\
Based on these studies, we adopt mass estimated from luminosity as a suitable indicator of group mass. 
%
%
\section{Large scale environment vs infall}
\subsection{Velocity field anisotropies}
In a first approximation, the mean velocity field around groups of galaxies can be suitably described by the spherical infall model. The velocities predicted by this model are in qualitative agreement with our results (see figure \ref{1_6} where the amplitude of the mean infall velocity field is strongly influenced by the group total mass). 
This model has also been successful in constraining group masses \citep{pivato2006} although we notice relevant departures from this simple picture given 
that the mass distribution around groups is strongly anisotropic due to the network of filaments, walls and clusters dominating the 
large scale Universe. For this reason the velocity fields surrounding groups exhibit a significant variance
which is strongly connected to surrounding large scale structures \citep{Kashibadze:2020,Courtois:2019,Tully:2019,Libeskind:2015}. 
These effects have been reported and analyzed in numerical simulations \citep{ceccarelli2011}
and in this work we aim at obtaining observational counterparts of these theoretical results.\\ 
 Notice that since only the LOS component of observational peculiar velocities are available,
the observational velocity field for groups with overdensities along the LOS 
can be strongly determined by galaxies residing in these filamentary regions.
On the other hand, groups located in global overdensities perpendicular to the LOS 
may have galaxies with velocity dominated by flows from underdense 
regions onto groups. 
In this section we examine these different cases by analysing the derived infall flow from over/under--dense regions by selecting two samples of groups with large scale surrounding regions parallel and perpendicular to the line of sight.
In order to perform this test we compute the density in a 
spherical region of radius $15$ Mpc h$^{-1}$ centered in the groups and consider separately the
regions close to the LOS, $ \theta < \frac{\pi}{6}$, and perpendicular to it
$ \frac{\pi}{3} < \theta < \frac{2\pi}{3}$.
This allows to separate the total subsample into two groups, those with predominant LOS overdensities ($\parallel$), 
and those predominant overdensities in the plane of the sky ($\perp$). The number of groups for each sample is stated in table \ref{table:samples}.\\
%
%
\begin{figure}
\centering
\includegraphics[width=1.0\linewidth,height=0.8\hsize]{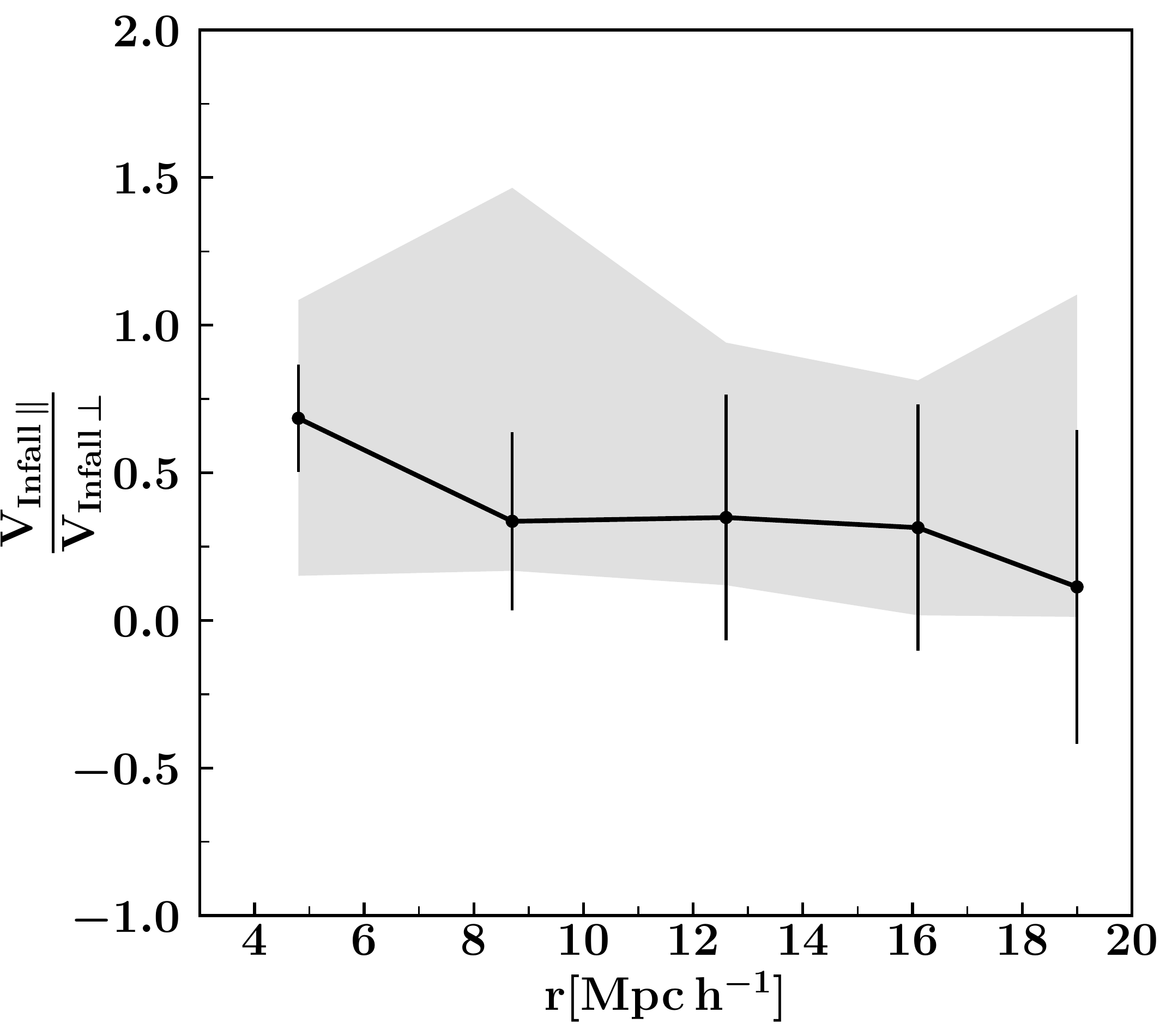}
 \caption{ 
Mean infall velocity ratio ($V_{infall \parallel}$ / $V_{infall\perp}$) for groups with overdensities in the direction parallel and perpendicular to the LOS in the observational data.
Error-bars indicate the uncertainty in the relative mean velocity and the gray shaded region shows the variance as in previous figures.
}
\label{1_8}
\end{figure}
With the aim at shedding light on the anisotropic infall onto groups in the observations, 
we estimate the mean velocity field around the two samples of groups.
Applying a similar analysis to that performed in the previous subsection we compute the 
ratio between both subsamples, and show the results in figure \ref{1_8}. 
As it can be noticed, mean velocities are significantly different as inferred from 
the KS test which gives a high statistical significance at more than the 5 $\sigma$ level.
The ratio  V{\footnotesize infall $\parallel$} / V{\footnotesize infall $\perp$}
for the two group samples remains below 1 (solid line, 
 V{\footnotesize infall $\parallel$}>V{\footnotesize infall $\parallel$}), 
indicating that the streaming motion of galaxies 
onto groups with large underdense regions is faster than those of galaxies from global overdensities.
This result is in sync with \citet{Pereyra:2019, Mahajan:2012, ceccarelli2011}, which showed that a relatively high galaxy density 
in the infalling regions of groups promotes tidal interactions with neighboring galaxies, resulting in a smaller magnitude of the velocities along these high density regions.
We stress the fact that both sub-samples of groups have  similar mass, and luminosity 
distributions so that these differences should be owed to an 
environment difference.

\subsection{Overall density around groups}\label{subsec:overall_density}
\begin{figure}
\centering 
    \includegraphics[width=1.0\linewidth,height=0.8\hsize]{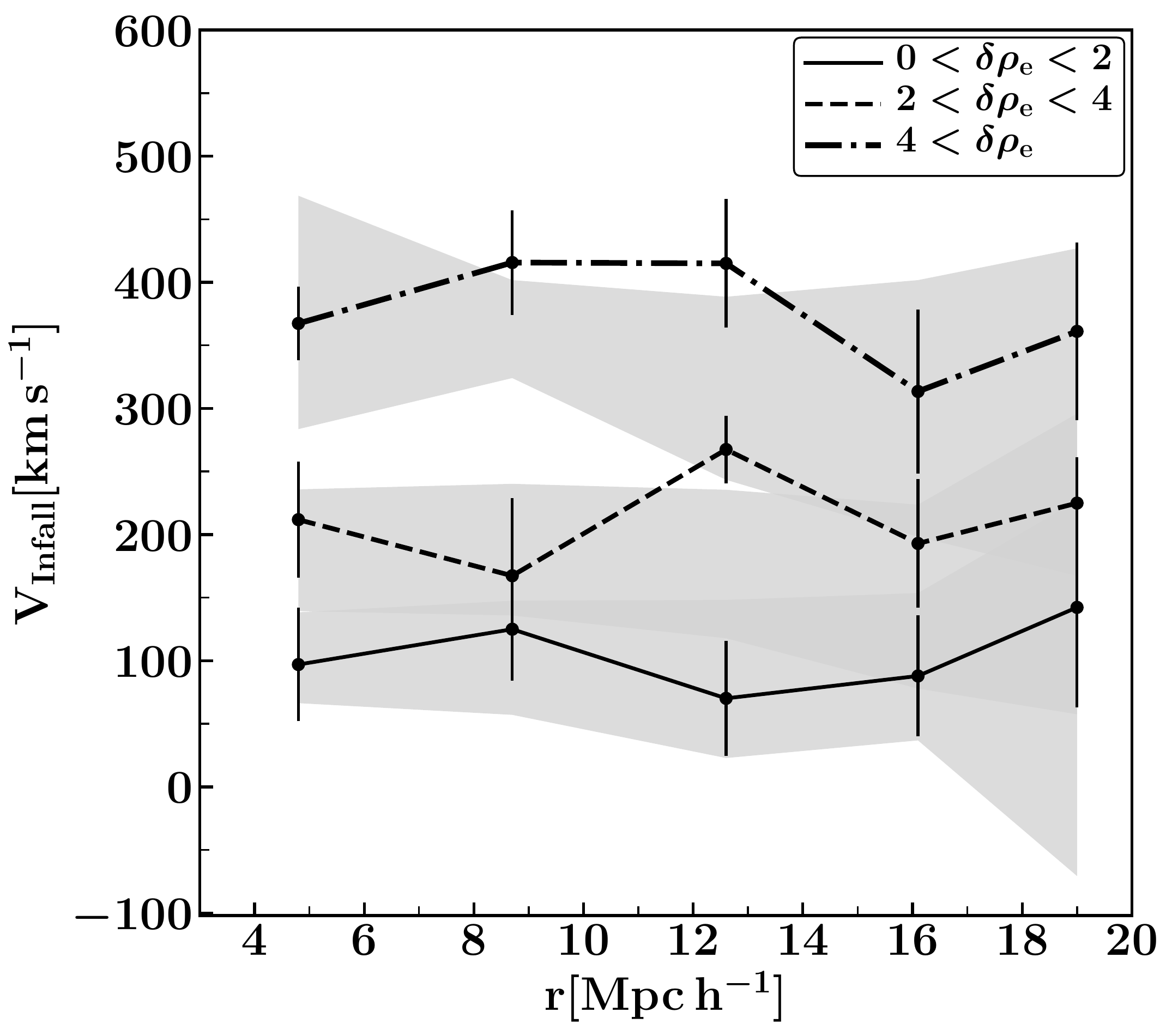}
\caption{
Dependence of infall onto groups from regions with different density environments
in observational data.  
The lines correspond to the mean infalling velocity onto groups embedded in regions with  
$\delta\rho_{e} > $ 4 (dot--dashed); 
2 $ < \delta\rho_{e} < $ 4 (dashed), and 0 $< \delta\rho_{e}< $ 2 (solid) 
The three gray shaded regions represent the variance from equivalent high, medium and low density environment in the mock catalogs respectively. 
Error--bar correspond to the uncertainties in the derived mean infalling velocities.   
 }
    \label{1_9a}
\end{figure}

The spherical infall model assumes an isotropic mass distribution of groups as well as their isolation so that infall pattern is associated to an isotropic convergent velocity field. Given that groups of galaxies are not isolated but immersed in the cosmic web, local irregularities can affect the local dynamics and even more, large scale 
bulk motions may imprint significant peculiar velocities to the galaxies beyond the small/intermediate effects associated to the galaxy groups \citep{Kitaura:2012,Hoffman:2018, Graziani:2019}.\\ 
We have analyzed the effects of the surrounding density field around groups on the infalling velocity pattern by selecting three subsamples of groups embedded in under/intermediate/over dense regions. 
For the environment characterization, we have computed the integrated galaxy overdensity on a region from $5$ Mpc h$^{-1}$ to $10$ Mpc h$^{-1}$ ($\delta\rho_e$) centered in the group and we have estimated the mean velocities as a function of group--centric distance for each sample.\\ 
The low density subsample is defined by 0 $<\delta\rho_{e} <$ 2,
the intermediate density subsample, by 2 $ < \delta\rho_{e} < $ 4,  and  
the high density subsample, by $\delta\rho_{e} >$  4.
\\
In figure \ref{1_9a} we show the resulting infall velocity pattern derived from the three group subsamples that consider their mean surrounding density. 
Solid line corresponds to 0 $<\delta\rho_{e} < $ 2, dashed line to 2 $ < \delta\rho_{e} < $ 4 and dot--dashed line to $\delta\rho_{e} > $ 4).\\ 
As it is can be seen in this figure, groups in all density environments show systematic infalling 
velocities over all range of distances explored with increasing amplitude for higher density environments.
In figure \ref{1_9a} it can be appreciated that as the density environment around groups increases, the amplitude of the streaming infall velocity field increases as well.
The KS tests confirm that the differences among infall velocity amplitudes of high, intermediate and low density environment
are statistically significant at a very high confidence level ($>$ 3 $\sigma$ for all samples).
As expected, we find a strong dependence of the velocity field around groups on both group mass and surrounding mass density at significantly large scales \citep{Einasto05, Einasto03a}.
\\
%
%
%
%
\section{Local vs global velocity fields}
Several works \citep[e.g.][]{Einasto03a,Einasto03b, Einasto05, Lietzen:2012} have shown that 
groups and clusters of galaxies in high density regions are richer, more massive, and more luminous than groups and clusters 
of galaxies in low density regions. 
In this context, the results obtained in the previous section could be biased by the inclusion of samples of different mass groups.
Therefore, in this section we examine the inferred mean infall pattern considering separately the effects of group mass and environment.

\subsection{Density around groups}
\begin{figure}
\centering
 \includegraphics[width=1.0\linewidth,height=0.8\hsize]{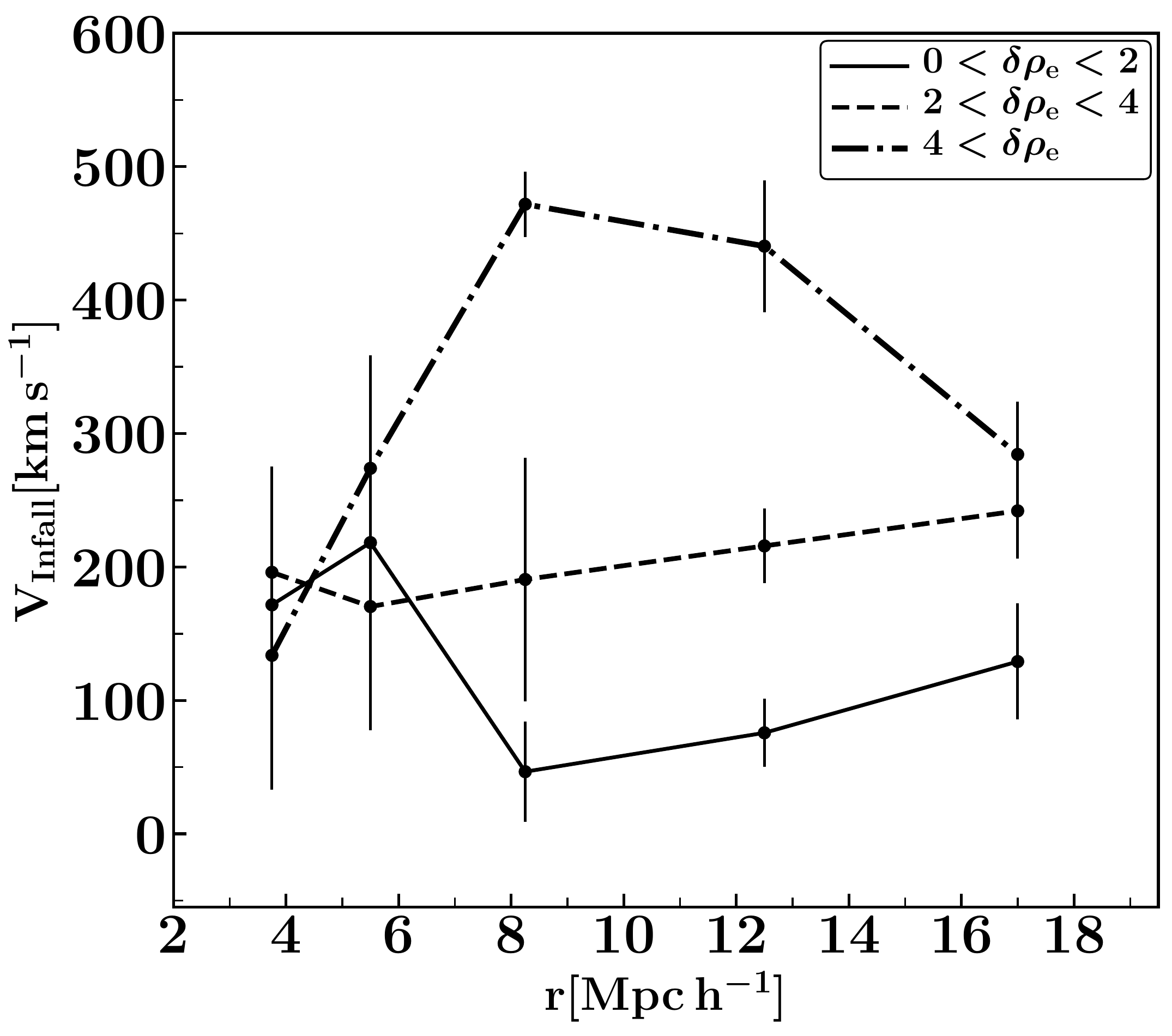}
\caption{
Same as figure \ref{1_9a} with observational group subsamples restricted to have a similar mass distribution.  
The line types correspond to the same environmental density ranges analyzed in figure \ref{1_9a} .
Error--bar indicate the uncertainty in the mean infalling velocities.   
 }
\label{1_9}
\end{figure}

We restrict the sample construction so that the groups have comparable statistical properties 
associated to mass. Given that we select samples of groups with similar statistical properties, 
the possible differences of their velocity fields should be associated only to environment.\\
We select three subsamples of groups having similar mass (and luminosity) distributions and 
embedded in regions of different overall galaxy density bins, where these density bins correspond to 
those 
in subsection \ref{subsec:overall_density}. 
Since the mass distribution of the three subsamples are similar, 
we expect similar contributions to the mean infall by the group itself. Therefore, the differences 
between the resulting velocity field of the two samples can be associated to the difference imposed by the surrounding environment.  
In figure \ref{1_9} we show the mean infall amplitude as a function of group--centric distance for groups 
in different galaxy density environments, solid lines $0 <\delta\rho_{e} < $ 2, dashed 
line $2 <\delta\rho_{e} <4$ and dot--dashed line $\delta\rho_{e} > $ 4. 
As it can be seen in the figure
the velocities are comparable for distances to the group centre smaller than $\sim$ 6 Mpc h$^{-1}$
 where the infall amplitude rises to V{\footnotesize infall} $\approx$ 200 km s$^{-1}$ at $r\approx$6 Mpc h$^{-1}$.  
This is consistent with the fact that the three samples have a comparable mass distribution. 
So, the effect of similar group masses 
on the surrounding velocity field can be clearly appreciated. 
On the other hand, in figure \ref{1_9} it can be noticed that the velocity curves differ significantly at large separations  ($r > 8$ Mpc h$^{-1}$), with associated KS test p--values < 0.01.
These results show the impact of the large scale environment
on the velocity field, in agreement with previous works \citep{Einasto05, Lietzen:2012}.
%
%

\subsection{Group mass dependence}
\begin{figure}
\centering
 \includegraphics[width=1.0\linewidth,height=0.8\hsize]{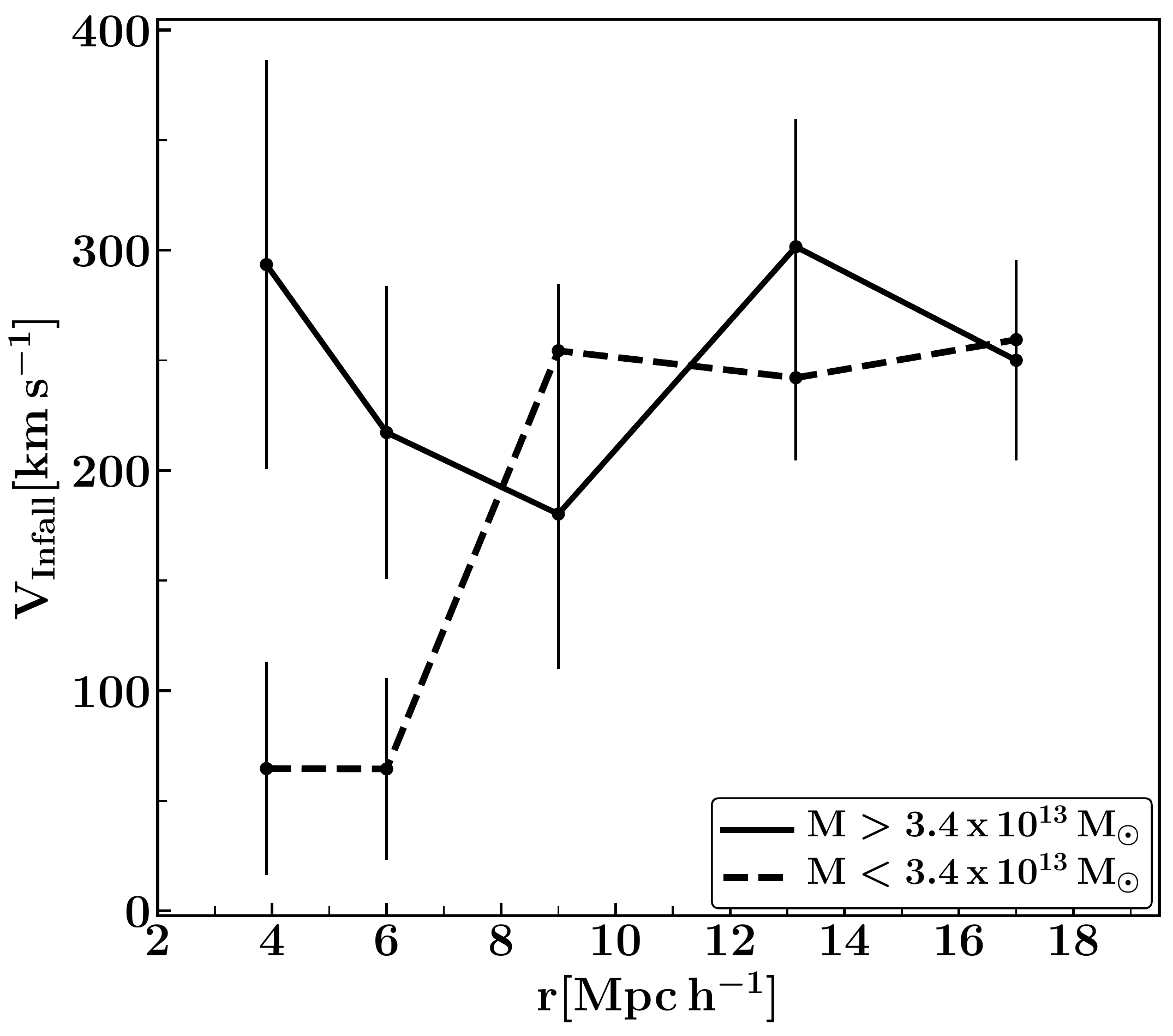}
\caption{Infall dependence on group mass in the observational data. The group subsamples are restricted to reside in similar density environments (1 $ < \delta\rho_{e} < $ 2.5).
The solid (shaded) line corresponds to high (low) mass groups (M larger (smaller) than M{\scriptsize median}. Error--bars represent the uncertainty in the derived mean
infalling velocities.}
\label{1_10}
\end{figure}

Once examined the effect of the global overdensity on the velocity field around 
poor galaxy groups, here we explore the contribution of group mass to the velocity field in an environment controlled subsample
We restrict the sample of groups to those residing in global
environments within a restricted galaxy overdensity (1 $ < \delta\rho_{e} < $ 2.5) corresponding to the 25 and 75 percentiles and comprises the 50\% of the total group sample. Within this constraint, we select two subsamples of groups with mass larger/smaller than the total sample median mass ($\approx$ 3.4 $\times$ 10$^{13}$ M$_{\odot}$). 
The mean/median mass for the larger (smaller) subsample is 10.4$\,$ x$\,$10$^{13}$/6.74$\,$x$\,$10$^{13}$ M$_\odot$ (18.8$\,$x$\,$10$^{12}$/17.8$\,$x$\,$10$^{12}$ M$_\odot$), with mean/median mass ratios 5.5/3.7.\\ 
Given that we select the two subsamples of groups in similar large--scale environments,
any differences on the peculiar velocity field should be attribute to the group mass.\\
We derive the mean infall amplitude for the two subsamples and show the results in
figure \ref{1_10}, where solid (dashed) line corresponds to group masses 
larger (lower) than the median.
As it can be noticed in this figure,   
high and low mass groups exhibit mean infall amplitudes clearly distinguishable closer to the group (r $<$ 8 Mpc h$^{-1}$):  
infall amplitude reaches 300 kms$^{-1}$ 
for high mass groups (solid line) 
and remains bellow 100 kms$^{-1}$ for low mass groups (dashed line)
at the smallest group--centric distances analyzed ($\sim$ 4 Mpc h$^{-1}$). 
The KS test shows that, up to  8 Mpc h$^{-1}$, 
the differences between the infall velocity amplitudes of these samples are statistically significant at the 2.5 $\sigma$ level.
Consequently, the infall amplitude ratio of  high--mass to low--mass groups 
is a factor 4.5 $\pm$ 1.0 in this close surroundings of groups , consistent with the mass ratio values 
(5.5 mean, 3.7 median) of the two group subsamples. 
This ratio is in very good agreement with the linear infall model predictions for the velocity field.\\
As it can be noticed in the figures \ref{1_9} and \ref{1_10} we can easily distinguish two regimes characterizing the peculiar velocity field at small and large scales.   

%
%
%
%
\section{Discussion}

Galaxy peculiar velocities originate in departures of motions from a pure Hubble flow induced by irregularities of the mass distribution at large scales. These motions have a statistical imprint in the redshift--space distortions of the two point correlation function and provide useful information on the global dynamics.
However, peculiar velocities can be directly estimated through redshift independent distance indicators and may be used to extract useful information on the dynamics of galaxies and galaxy systems besides the correlation function analysis.
Since peculiar velocity measurements provide a direct trace of the mass distribution
 there are no issues regarding galaxy bias which add complexity in redshift--space correlation analysis. For this reason, direct estimates of galaxy peculiar velocities may provide useful cosmological test in forthcoming surveys with more accurate determinations in the local Universe.
In this work we have focused our analysis on the measurements of infall of galaxies onto groups, and its dependence on group mass and environment.
Since groups of galaxies are part of the cosmic web, the local dynamics cannot be simply addressed through a spherical infall model given that large scale 
bulk motions affect the local effects associated to group mass overdensities.

Firstly we apply the spherical infall model to derive the mean infall velocity pattern onto groups of different characteristics. Since  only the line--of--sight projection of galaxy peculiar velocities can be obtained, the derived infall velocity field relates to galaxy positions relative to both the group center and the observer (see equation \ref{coseno}).
Present data are affected by large distance measurement uncertainties which lead to large peculiar velocity errors. This fact is associated to the presence of systematic effects of distance estimates that can bias the inferred velocity field.
To properly address the impact of distance measurement uncertainties on 
the mean infall mean velocity determination by assigning suitable errors to mock catalogs constructed similar to observations\\ 
We obtain accurate determinations of the mean infall velocity profile around groups of different mass range and in different environments which exhibit amplitudes in the range $200$ to $ 350$ km/s, entirely consistent with numerical simulation results.\\
We have extended our work by considering the impact of 
large--scale environment on the mean infall galaxy velocity onto groups.\\
We compare the effects of large--scale inhomogeneities around 
galaxy groups on the mean infall velocity, finding these effect significantly smaller along the direction of high density enhancement
compared to those along low density. This provides evidence that groups grow in a different fashion from filaments  than elsewhere.
These results are in agreement with previous studies of the environmental effects on groups and clusters growing by mass accreted from their surrounding structures ranging from isolated galaxies to large groups which
merge to form larger systems \citep{mcgee2009}.
\citet{mcgee2009} results also show that a large fraction of galaxies accreted onto clusters were formerly in groups.\\ 
We obtain a significant dependence of the mean infall pattern on 
the group large--scale environment consistent with 
higher infall velocities onto groups residing in large overdense 
regions.\\
We recall that infall models are mainly based on assumptions of spherical 
symmetry onto an isolated mass overdensity. This simplified scenario
may have an impact on total group mass determinations taking into account our
studies of
the streaming motions dependence on global density environment and 
anisotropic mass distribution around groups in observations.\\
Due to upcoming improvement on  
observational data and peculiar velocity precision, studies such as those presented in this work may provide 
aid in our current understanding of the
growth of structures in the Universe.

\begin{acknowledgements}
This work has been partially supported by Consejo de Investigaciones 
Cient\'{\i}ficas y T\'ecnicas de la Rep\'ublica Argentina (CONICET), and the
Secretar\'{\i}a de Ciencia y T\'ecnica de la Universidad Nacional de C\'ordoba
(SeCyT).
\end{acknowledgements}

%
%

\bibliographystyle{aa} 
\bibliography{infall.bib} 
\end{document}